\documentclass[a4paper,11pt]{article}
\usepackage{pos}
\usepackage{multirow}
\usepackage{booktabs}

\DeclareMathOperator\erf{erf}

\title{$^{17}$O enrichment of CaWO$_4$ crystals for spin-dependent DM search }
\ShortTitle{$^{17}$O enrichment of CaWO$_4$ crystals for spin-dependent DM search }

\author*[a]{Angelina Kinast}
\author[a,b,c]{Andreas Erb}
\author[a]{Stefan Schönert}
\author[a]{Raimund Strauss}
\author[c]{Jürgen Haase}

\affiliation[a]{Physik-Department, Technische Universit\"at M\"unchen,\\ James-Franck-Straße 1,D-85747 Garching, Germany}

\affiliation[b]{Walther-Mei\ss ner-Institut f\"ur Tieftemperaturforschung,\\ Walther-Mei\ss ner-Straße 8, D-85748 Garching, Germany}

\affiliation[c]{Felix Bloch Institute for Solid State Physics, University of Leipzig,\\ Linn\'estraße 5,  D-04103 Leipzig, Germany
}

\emailAdd{angelina.kinast@†um.de}

\abstract{
For many years, various experiments have attempted to shed light on the nature of dark matter (DM). This work investigates the possibility of using CaWO$_4$ crystals for the direct search of spin-dependent DM interactions using the isotope $^{17}$O with a nuclear spin of 5/2. Due to the low natural abundance of 0.038$\%$, an enrichment of the CaWO$_4$ crystals with $^{17}$O is developed during the crystal production process at the Technical University of Munich. Three CaWO$_4$ crystals were enriched, and their $^{17}$O content was measured by nuclear magnetic resonance spectroscopy at the University of Leipzig. This paper presents the concept and first results of the $^{17}$O enrichment and discusses the possibility of using enriched crystals to increase the sensitivity for the spin-dependent DM search with CRESST.}

\FullConference{ XVIII International Conference on Topics in Astroparticle and Underground Physics (TAUP2023)\\
 28.08-01.09.2023\\
University of Vienna\\}

\begin{document}
\maketitle

\section{Introduction}

The nature of dark matter (DM) is perhaps one of the greatest mysteries in modern cosmology and astroparticle physics. Despite numerous observations pointing to the existence of DM, it has not yet been detected.
DM direct detection experiments aim to measure the direct scattering of DM particles with Standard Model particles. This interaction can be either spin-independent or spin-dependent. In the spin-independent case, the expected rate scales with the mass number squared (A$^2$), whereas, in the spin-dependent case, it scales with the nucleon spin coefficients ( $\langle$ S$_{p/n}$ $\rangle ^2$). Therefore, only isotopes with a non-vanishing nuclear spin, such as $^{17}$O with an unpaired neutron spin of I = 5/2, can be used for spin-dependent searches. 
The CaWO$_4$ target crystals for the CRESST DM experiment have been grown in-house at the Technical University of Munich (TUM) for many years, focusing on high radiopurity, good optical quality and stress-minimised crystal lattice \cite{erb2013growth,kinast2023characterisation}. CaWO$_4$ detectors reach thresholds as low as 30.1\,eV \cite{detA} and can be used for spin-independent and spin-dependent DM searches. However, $^{17}$O has a natural abundance of only 0.038\,$\%$, which limits the sensitivity for spin-dependent DM searches. Therefore, enriching CaWO$_4$ with $^{17}$O to, e.g. 3.8\%, so by a factor of 100, would increase the sensitivity for spin-dependent DM searches by two orders of magnitude.

\section{$^{17}$O enrichment of CaWO$_4$ }
The oxygen diffusion enrichment procedure is based on the experience with CaWO$_4$ crystal growth at TUM. After the growth, the CaWO$_4$ crystals are annealed to remove oxygen vacancies formed during the growth. 
For this, the crystals are placed in a 100\,$\%$ oxygen atmosphere  at 1400\,$^{\circ}$C for 20\,h. At this temperature, oxygen diffuses into the crystal lattice, filling the vacancies. This diffusion is driven by the gradient formed by the vacancies. Afterwards, the crystals are cooled to 1200\,$^{\circ}$C with 5\,K/h and to room temperature with 50\,K/h. 

The annealing process cannot be used for enrichment since the oxygen gas is lost in the procedure, making it impractical with $^{17}$O enriched oxygen gas. In addition, enrichment is unnecessary for crystal parts not used in detector production, including the tail, shoulder and outer layers.  Therefore, the furnace was adapted by adding an airtight quartz glass tube partially embedded in the furnace and connected to a gas handling system outside the furnace. The annealed and fully prepared detector crystals were placed in this tube for enrichment. The tube is then filled with 70$\%$ $^{17}$O enriched oxygen gas and heated to 1300\,$^{\circ}$C, the highest temperature that the quartz glass tolerates. At this temperature, oxygen diffuses between the gas and the crystal without the assistance of a gradient. The CaWO$_4$ diffusion coefficient for this process is unknown and depends on chemical bonding and temperature. For other minerals, it ranges over several orders of magnitude. Hence, this work investigates whether it is possible to enrich CaWO$_4$ with $^{17}$O by means of this process. For this study, three detector crystals (C1, C2 and C3) were used. C1 and C2 have dimensions of 20x20x10mm³, while C3 is comparatively smaller at 5x5x5mm³. The crystals were enriched for 20 hours (C1) and 200 hours (C2 and C3) before being cooled to room temperature at 50\,K/h. 

\begin{figure}[b!]
\centering
\includegraphics[width=0.90\linewidth]{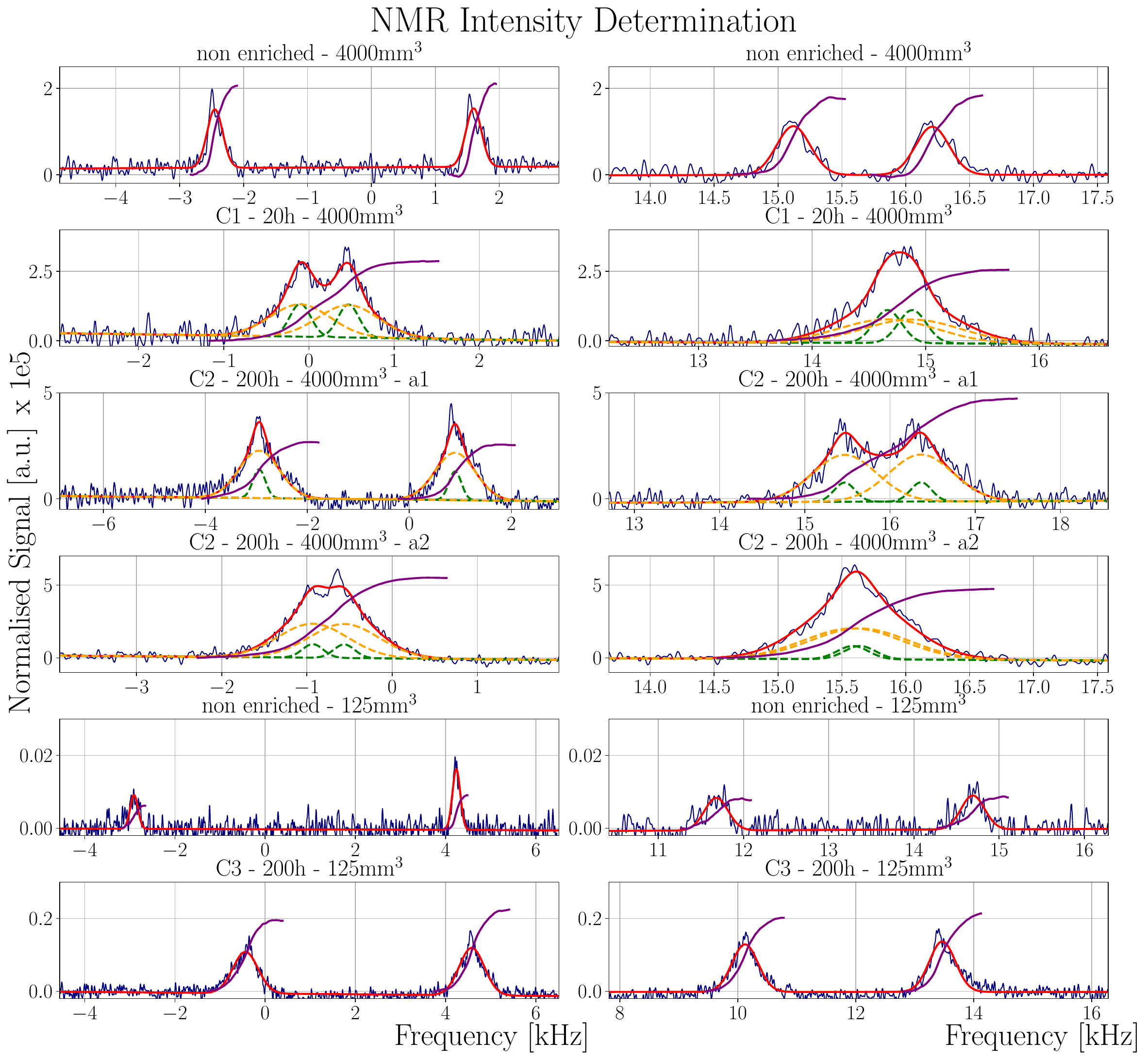}
\caption{NMR intensity determination for crystals C1, C2 and C3. Shown is the NMR signal normalised to the measurement time against the frequency deviation from the excitation frequency of 67.8\,MHz. All four peaks are fitted with a Gaussian each, in case of overlapping Gaussians by a combined Gauss model (see red, orange and green fit curves). Additionally, a numerical integration was performed. The cumulative integrals are shown by the purple curves. 
}
\label{NMRFits}
\end{figure}
\section{Determination of the $^{17}$O content with NMR }
Nuclear Magnetic Resonance (NMR) spectroscopy of the $^{17}$O nuclear spin states can determine the amount of nuclei present in the sample. In NMR, the material is placed in a strong magnetic field, B$_0$, to polarise the nuclear moments along the field axis. In a subsequent resonance experiment, the resulting nuclear magnetisation is flipped into the plane perpendicular to the field (90$^{\circ}$ radio frequency pulse at the nuclear Larmor frequency, $\omega = ^{17}\gamma B_0$) and the precessing magnetisation is detected with a simple wire coil. The resulting transient voltage has an amplitude proportional to the number density of $^{17}$O nuclei and oscillates near the Larmor frequency. 

A Fourier transform of the transient reveals the spectral distribution of the nuclear resonances, which are influenced by the chemical structure and the inter-nuclear magnetic dipole interactions. For CaWO$_4$, the WO$_4$ tetrahedra are slightly distorted by Ca atoms, resulting in four slightly different oxygen positions. This could lead to four different NMR lines. In addition, since the $^{17}$O nucleus has a spin of 5/2, there is an electric hyperfine interaction of the nuclear quadrupole moment with the local electric field gradient. This quadrupole interaction can lead to 2I = 5 resonances for each oxygen nucleus, with a central transition and four symmetrical space satellites. Thus, we expect 4 x 5 = 20 resonances, with the quadrupole splitting probably exceeding the shift due to the slight chemical differences (we are not aware of any $^{17}$O NMR of CaWO$_4$).

This is indeed what we observe, and we focus on the four central lines. It should be noted that observing $^{17}$O NMR in natural abundance in solids is challenging for signal-to-noise reasons, and one has to rely on signal averaging. However, this is difficult in CaWO$_4$ because the nuclear relaxation time required to build up the nuclear polarisation is quite long (we estimate a T$_1$ of about three minutes). This weak nuclear relaxation is expected for clean crystals but prevents a comprehensive study of the full orientation dependence. 

Nevertheless, the intensities of the four central lines must be equal (equal distribution of $^{17}$O among the four tetrahedral positions), and their intensity increase proportionally to the enrichment factor. The inter-nuclear dipole interaction is very weak in the non-enriched materials since the nuclear moments of W and Ca are very weak compared to those of $^{17}$O. In the enriched sample, the enrichment is assumed to be stronger at the surface of the crystal, hence, we can expect narrow lines from the inner parts of the crystal and broader lines from the enriched regions. %Calculation of the associated linewidths is standard NMR procedure.

\begin{table}[b]
\centering
\begin{tabular}{lllll}
      
\toprule	
				& 					\multicolumn{3}{c}{Enrichment factors} 															&  $^{17}$O Abundance		\\
				&Gauss Fit							& Numeric Integral 						& Combined							& Combined				\\
\midrule
C1 (4000\,mm$^3$, 20h) 		& 3.22 $\pm$ 0.08						& 3.31 $\pm$ 0.15		  				&3.26 $\pm$ 0.10							&	(0.124 $\pm$ 0.004)\,$\%$ \\ 
\midrule
C2 (4000\,mm$^3$, 200h)  - a1	& 7.22$\pm$ 0.20						&  7.1 $\pm$ 0.7						& \multirow{2}{*}{7.26 $\pm$ 0.23} 			&	 \multirow{2}{*}{(0.276 $\pm$ 0.009)\,$\%$}  	\\ 
C2 (4000\,mm$^3$, 200h) - a2	& 7.22$\pm$ 0.20						&  7.5 $\pm$ 0.4						&									&		\\ 
\midrule
C3 (125\,mm$^3$, 200h) 		& 26.7 $\pm$ 1.5						& 27 $\pm$ 5		  				&27.1 $\pm$ 2.9							&	(1.03 $\pm$ 0.11)\,$\%$ \\ 
\bottomrule
    \end{tabular}
    \caption{Resulting enrichment factor determined by the Gauss fit, the numerical integral and a combination of both. The resulting average $^{17}$O abundance of each crystal is shown in the last column.   }
    \label{FitResultsEnrichmentFactor} 
\end{table}

To determine the $^{17}$O content, two samples of equal size – one enriched and one standard CaWO$_4$ – were measured at the NMR setup at Universität Leipzig and compared. For this analysis, we assessed the intensities by fitting a Gaussian model or by using numerical integration. Both methods were utilised and compared to reduce model-dependent uncertainties. 
Figure \ref{NMRFits} displays fits to the four major peaks around the $^{17}$O Larmor frequency. As the crystals were oriented differently, the peak positions shift between measurements, as evident in the two measurements of C2 along direction a1 relative to along direction a2. Enrichment factors were calculated by integrating all four peaks and comparing the result to the non-enriched sample. 
Due to a concentration gradient from the surface to the bulk there will be a gradient of the local magnetic field that the nuclei sense, causing broadening of the resonance. In the simplest approximation, the spectra were fitted by two Gaussians for each peak.
Table \ref{FitResultsEnrichmentFactor} shows the resulting factors for both reconstruction methods, which are in good agreement within their respective uncertainties. The measurements of C2 along two different axes demonstrate that the outcomes are not influenced by the crystal's orientation.

\section{Results and Outlook}
The NMR results show that CaWO$_4$ crystals can be successfully enriched with $^{17}$O by the method described above, although not by the amount desired to improve the sensitivity of CRESST. However, crystals C1 and C2 could improve the current CaWO$_4$ exclusion limits (see \cite{detA}) by a factor of 3.26 and 7.26, respectively. To gain insight into the necessary modifications needed for enrichment by a factor of 100, the diffusion coefficient in CaWO$_4$ at 1300\,$^{\circ}$C is calculated from the data. To estimate this, we utilise the Van-Ostrand-Dewey solution of the diffusion equation, as shown in Equation \ref{DiffusionSolution}.
Here, the diffusion can be expressed as

\begin{equation}
\frac{C_c - C_s}{C_0 - C_s}= \erf \left( \frac{x}{2 \sqrt{Dt}} \right)
\label{DiffusionSolution}
\end{equation}

with C$_0$, the concentration of $^{17}$O prior to the enrichment, C$_s$, the $^{17}$O gas concentration, and C$_c$, the concentration at a certain distance x from the surface. A leak occurred during the enrichment of C2 and C3, resulting in a larger uncertainty in the value of C$_s$. 
The resulting diffusion coefficients can be calculated assuming an isotropic diffusion coefficient of CaWO$_4$. The results are shown in table \ref{DiffusionCoefficients} together with an averaged diffusion coefficient from all three measurements. 

\begin{table}[h!]
\centering
\begin{tabular}{lllll}   
\toprule	
CaWO$_4$ crystal & C1 & C2 &C3 & Mean \\
\midrule	
Diffusion Coefficient [10$^{-16}$m$^2$/s] & 1.10 $\pm$ 0. 11 & 1.2 $\pm$ 0.4 & 2.1 $\pm$ 0.8 & 1.46 $\pm$ 0.31  \\ 
\bottomrule
    \end{tabular}
    \caption{Calculated diffusion coefficients for $^{17}$O diffusion in CaWO$_4$ at  1300\,$^{\circ}$C from all three crystals measured in this work. Additionally, the mean of these three values is stated.  }
    \label{DiffusionCoefficients} 
\end{table}

With this coefficient, the $^{17}$O distribution can be calculated for any enrichment time as a function of distance from the surface or the average enrichment. The results show that only the upper tens of micrometres are enriched by this method, and due to the square root dependence of time in the equation \ref{DiffusionSolution}, extending the enrichment time would not result in significantly larger enrichment factors. The same applies to an increase in the diffusion coefficient. Based on the temperature dependency of other minerals in \cite{farver2010oxygen}, an increase of the diffusion coefficient by approximately 2.7 ± 0.3 was estimated for CaWO$_4$ when the temperature is increased to 1400\,$^{\circ}$C. Hence, increasing temperature would not considerably enhance the enrichment factors.  
A method to accomplish the targeted enrichment factor of 100 is to enrich the CaWO$_4$ powder before growth since the powder's grain size is typically in the range of a few tens of micrometres, similar to the measured diffusion depth in this study. Hence, powder enrichment presents a viable alternative to crystal enrichment. The enrichment facility at TUM is currently being upgraded to evaluate the effectiveness of this method.

\acknowledgments
This work has been funded by the Deutsche Forschungsgemeinschaft (DFG, German Research Foundation) under Germany's Excellence Strategy – EXC 2094 – 390783311 and through the Sonderforschungsbereich (Collaborative Research Center) SFB1258 ‘Neutrinos and Dark Matter in Astro- and Particle Physics’ and by the BMBF 05A20WO1.

%\begin{thebibliography}{99}

\bibliographystyle{JHEP}
\bibliography{skeleton.bib}

%\bibitem{...}
%....

%\end{thebibliography}

\end{document}